\documentclass[useAMS,usenatbib]{mn2e}
\usepackage{psfrag,graphicx}
\usepackage{txfonts}
\usepackage{natbib}
\usepackage{ulem}
\usepackage{color}
\usepackage{definitions}
\usepackage{xspace}

\title[Impact of freeze-out on collapsing clouds]{The impact of freeze-out on collapsing molecular clouds}
\author[S. Hocuk, S. Cazaux and M. Spaans]{S. Hocuk$^{1}$\thanks{E-mail: seyit@astro.rug.nl}, S. Cazaux$^{1}$ and M. Spaans$^{1}$\\
$^{1}$Kapteyn Astronomical Institute, University of Groningen, P. O. Box 800, 9700 AV Groningen, Netherlands}

\begin{document}

\date{Submitted 9 October 2013; Accepted 31 October 2013}
\maketitle

\label{firstpage}
\begin{abstract}
{Atoms and molecules, and in particular CO, are important coolants during the evolution of interstellar star-forming gas clouds. The presence of dust grains, which allow many chemical reactions to occur on their surfaces, strongly impacts the chemical composition of a cloud. At low temperatures, dust grains can lock-up species from the gas phase which freeze out and form ices. In this sense, dust can deplete important coolants. Our aim is to understand the effects of freeze-out on the thermal balance and the evolution of a gravitationally bound molecular cloud. For this purpose, we perform 3D hydrodynamical simulations with the adaptive mesh code \textit{FLASH}. We simulate a gravitationally unstable cloud under two different conditions, with and without grain surface chemistry. We let the cloud evolve until one free-fall time is reached and track the thermal evolution and the abundances of species during this time. We see that at a number density of 10$^4$\cmm3 most of the CO molecules are frozen on dust grains in the run with grain surface chemistry, thereby depriving the most important coolant. As a consequence, we find that the temperature of the gas rises up to $\sim$25\K. The temperature drops once again due to gas-grain collisional cooling when the density reaches a few$\times$10$^4$\cmm3. We conclude that grain surface chemistry not only affects the chemical abundances in the gas phase, but also leaves a distinct imprint in the thermal evolution that impacts the fragmentation of a star-forming cloud. As a final step, we present the equation of state of a collapsing molecular cloud that has grain surface chemistry included.}
\end{abstract}

\section{Introduction}
\label{sec:introduction}
Star-forming gas clouds that are optically thin can easily radiate away their internal heat during the course of their lifetimes. Radiation is emitted through (ro-vibrational) line transitions of species in the gas phase (n $\lesssim 10^4$\cmm3) and/or by black body radiation from dust grains (n $\gtrsim 10^4$\cmm3). The cooling through these processes is very efficient and poses a principle problem to star formation models which predict that most stars should form within a free-fall ($t_{\rm ff}$) time. This contradicts observations which estimates the star formation rate to be $\sim$1\% per free-fall time \citep{2012ApJ...745...69K} and an efficiency of about 5\% \citep{2011MNRAS.415.3439D, 2011ApJ...729..133M}. Numerical studies that include stellar feedback such as winds, outflows, or radiative feedback \citep{2009MNRAS.392.1363B, 2010ApJ...713.1120K} have made varying successes in obtaining more realistic star formation efficiencies. However, direct impact on the efficiency through the depletion of coolants is never before fully investigated.

Dust particles in the interstellar medium (ISM) enhance molecular abundances by acting as a catalyst for the formation of molecules. Their involvement in chemical reactions affects the thermodynamic properties of a molecular cloud and, henceforth, its whole evolution \citep{2010A&A...522A..74C, 2012A&A...537A.102M}. Dust chemistry can, however, also have an adverse effect on molecules due to its dual nature. Apart from catalyzing the formation of efficient coolants, releasing them into the gas phase and aiding (the early stages of) gravitational collapse, dust particles can also lock-up important coolants by freezing species on their surfaces \citep{2001ApJ...557..736G} thereby changing the equation of state (EOS) and, as a consequence, hinder cloud contraction and fragmentation.

Depletion of gaseous CO has been observed in many regions such as low-mass starless cores and high-mass star-forming infrared dark clouds \citep{2002ApJ...569..815T, 2004A&A...416..191T, 2011ApJ...738...11H, 2012MNRAS.423.2342F}. At densities exceeding $\sim$3$\times10^4$, depletion is not expected to have a significant impact on the thermal balance since gas and grain temperatures will be tightly coupled. However, below these densities depletion can impact the thermal balance by decreasing the cooling which in turn increases the temperature if the gas, while the low gas-grain coupling will not affect the dust temperature. Observational data indicates that depletion factors of upto 80 and gas temperatures of upto 25\,K can exist in high-mass pre-stellar cores \citep{2012MNRAS.423.2342F}. This might leave the warmer gas to coexist with the colder dust when there is sufficient time for depletion to become significant \citep{2001ApJ...557..736G}. The interplay between gas and dust is therefore an important physical factor to take into account in simulations of molecular cloud formation and evolution. 

In this study, we show with numerical simulations how strongly a collapsing cloud is affected due to freeze-out by including detailed gas and grain surface chemistry (GSC) in our models and discuss the impact of freeze-out on cloud evolution.

\section{Numerical method}
\label{sec:numericalmethod}
\subsection{The simulation code}
The simulations in this work have been performed using the hydrodynamical code \textit{FLASH}4 \citep{2000ApJS..131..273F, Dubey2009512}. The adaptive mesh refinement (AMR) code \textit{FLASH} is well-suited to handle the type of calculations for this study. Our work captures the physics that act on small, micro sizes and on large, (sub-)parsec scales. 

Our code is equipped with hydrodynamics, chemistry, thermodynamics (using heating and cooling rates), turbulence, multi-species, gravity, column densities (for UV extinction), and adaptive refinement criteria based on Jeans length. Many of the modules are well-tested and are either provided by \textit{FLASH} or found in earlier works of \cite{2010A&A...522A..24H, 2011A&A...536A..41H}. The key physics modules, thermodynamics and chemistry, are discussed in Section \ref{sec:analyticalmethod}.

\subsection{Initial conditions}
We simulate a spherical molecular cloud with an initial number density of \nhtot = $10^{3}$\cmm3 and an initial temperature of 10\K in order to follow its chemistry and temperature during collapse. \nhtot is the total hydrogen nuclei number density. The cloud has a radius of 4.25\pc and a total mass of 7250 \msol. We place our cloud in a 3D box of size 15$^3$\pc with a number density that is typical for the ISM of \nhtot = 1\cmm3, such that the total mass outside of the cloud is negligible relative to the cloud mass. The simulation box has periodic boundary conditions for gravity and space.

The molecular cloud is a gravitationally bound system and is gravitationally unstable. We initiate our molecular cloud with turbulent conditions that are typical for the Milky Way, $\rm \sigma_{turb}$ = 1\kmps, and apply this over all scales with a power spectrum of $\rm P(k) \propto k^{-4}$, following the empirical laws for compressible fluids \citep{1981MNRAS.194..809L, 1999ApJ...522L.141M, 2004ApJ...615L..45H}. The turbulence in this work is not driven and is not strong enough to support gravitational collapse.

We place our molecular cloud in a star-forming region with a background UV radiation flux of 1\,$G_{0}$ in terms of the Habing field \citep{1968BAN....19..421H}. This agrees with the -on average- ISM conditions of our Milky Way. This creates a temperature gradient that ranges from 10\K inside the cloud to 50\K outside, unattenuated regions \citep{2005A&A...436..397M}. The isothermal sound speed of the cloud in this case ranges between $\rm c_{s}=0.20$\kmps to 0.45\kmps.

We refine our grid according to a Jeans length criterion, with a resolution of 12 cells per Jeans length. Given a maximum resolution of 256$^3$ cells the spatial resolution yields $5.7\times10^{-2}$\pc.

\section{Analytical method}
\label{sec:analyticalmethod}
\subsection{Time-dependent chemistry}
To determine the chemical composition of the gas, as well as the species frozen out onto dust, we use time-dependent rate equations that include both gas and surface reactions. In our models, we include 32 species from which 23 species are in the gas phase and 9 species are in the solid phase on dust grains. We assume a dust-to-gas mass ratio of 0.01 for Solar metallicity and adopt the grain-size distribution of \cite{2001ApJ...548..296W}. Our selection of species are: H, H$^{+}$, H$^{-}$, H$_{2}$, H$_3^{+}$, C, C$^{+}$, C$^{-}$, O, O$^{+}$, O$^{-}$, O$_{2}$, CO, CO$_{2}$, OH, OH$^{+}$, H$_{2}$O, H$_{2}$O$^+$, H$_{3}$O$^+$, HCO, HCO$^+$, H$_{2}$CO, e$^{-}$, $\bot$H, $\bot$H$_{2}$, $\bot$C, $\bot$O, $\bot$O$_{2}$, $\bot$OH, $\bot$CO, $\bot$H$_{2}$O, $\bot$HCO. The symbol $\bot$ denotes a bound/ice species.

The species can react with one another and we included all known important reactions involving our chosen species. In total, we have 169 chemical reactions. The gas phase reactions are taken from the Kinetic Database for Astrochemistry \citep[\textit{KiDA},][]{2012ApJS..199...21W}. We refer the reader to the \textit{KiDA} database for these equations and their corresponding parameters. 

The reactions on grain surfaces are obtained from \cite{2010A&A...522A..74C} and comprise 50 reactions. Chemical reactions involving dust grains constitute five different physical mechanisms, i.e.,
\begin{description}
 \item[(A)] accretion of gas phase species by grains,
 \item[(B)] evaporation of bound/ice species,
 \item[(C)] two-body reactions on grain surfaces,
 \item[(D)] direct cosmic-ray processes,
 \item[(E)] photo-processes that include photodissociation, photodesorption, and cosmic-ray induced photo-processes.
\end{description}
The details of these reactions and their implementations are further explained in \cite{2010A&A...522A..74C}. In this study we focus on the effect of freeze-out of species on the thermal evolution of a cloud. The two-body reactions on surfaces do not play a major role in our simulations. 

The initial conditions considered in our simulations are similar to conditions of translucent clouds by \cite{2009ApJ...690.1497H}. In order to have realistic initial conditions our model without GSC follows the low extinction (A$_v \sim 3$) values of \cite{2009ApJ...690.1497H} when there are no ices, i.e., 
$[H_2] = 0.5$, 
$[HI] \simeq 5\times 10^{-5}$, 
$[CO] \simeq 1.3\times 10^{-4}$, 
$[O] \simeq 1.3\times 10^{-4}$, 
$[H_2O] \simeq 5\times 10^{-7}$, 
and the translucent cloud conditions (A$_v \sim 5$) when ices are present for our model with GSC, i.e., 
$[H_2] = 0.5$, 
$[HI] \simeq 5\times 10^{-5}$, 
$[CO] \simeq \times 10^{-5}$, 
$[O] \simeq \times 10^{-6}$, 
$[H_2O] \simeq \times 10^{-7}$, 
$\bot [CO] \simeq 5\times 10^{-5}$, and 
$\bot [H_2O] \simeq 2\times 10^{-4}$.
These final values were obtained by running a low resolution run without gravity for 1 Myr until a convergence in abundances was reached.

\subsection{Thermal processes}
To properly solve the thermal balance and obtain our final gas and dust temperatures, we included several radiative and collisional heating and cooling functions. The considered 6 heating terms are; photo-electric emission, \hh photodissociation heating, \hh collisional de-excitation heating, cosmic ray heating, gas-grain collisional heating (also a cooling term), and compressional heating. The heating functions and rates are obtained from \cite{2005A&A...436..397M, 2009A&A...501..383W}. 

For our cooling rates, we consider the following 6 terms; electron recombination with PAHs cooling, electron impact with H (Ly-alpha cooling), metastable transition [OI]-630 nm cooling, fine-structure line cooling for [OI]-63 $\mu$m \& [CII]-158 $\mu$m, gas-grain collisional cooling (also a heating term), and molecular cooling by \hh, \co, \oh, and \h2o. The cooling functions and rates are obtained from \cite{1993ApJ...418..263N, 1995ApJS..100..132N, 2005A&A...436..397M, 2009A&A...501..383W, 2010ApJ...722.1793O}.

\section{Results}
\label{sec:results}
We analyze the simulation output at the moment the cloud has reached its assumed free-fall time of \tff = $1.63\times10^{6}$\yr. Both simulations reach a density of n = 10$^7$\cmm3 around the same epoch.

\subsection{Phase diagrams}
\label{sec:phasediagrams}
\begin{figure}
\includegraphics[scale=0.455]{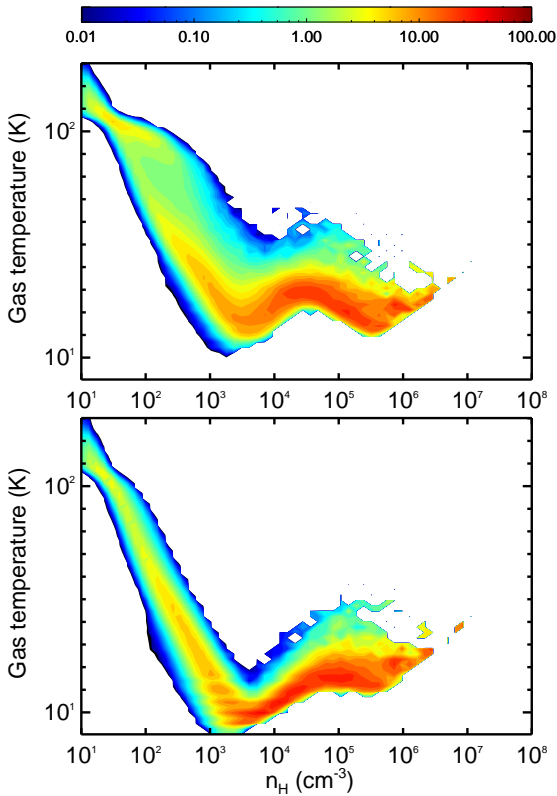}
\caption{Temperature-density phase diagram. The top panel displays the phase diagram of the model with GSC. The bottom panel shows the phase diagram of the model without GSC. Both images are displayed at $t = t_{\rm ff}$. The color demonstrates the amount of mass lying in a contoured region, ranging from $0.01$ \msol (blue) to $100$ \msol (red).}
\label{fig:phases}
\end{figure}

The gas temperature of a molecular cloud will vary strongly with its number density, since the heating and cooling rates depend on density and chemical abundance. In Fig. \ref{fig:phases}, we show the temperature evolution of our two cloud models. 
One can clearly see that the model with grain surface chemistry displays a different evolution in temperature than the model with only gas phase chemistry. There is a temperature bump between a density of 10$^4$\cmm3 -- 10$^5$\cmm3 in the GSC model. 

At densities of 10$^3$ to 10$^4$\cmm3 most of the mass is lying at a temperature of around 15\K in the surface chemistry model, whereas in the pure gas chemistry model this is around 10\K. The gas cannot cool as efficiently in the GSC model due to the freeze-out of CO at low ($\rm T < 20$\K) temperatures and thus the depletion of the main coolant. Following the depletion of CO from the gas, the temperature increases to 20--30\K until a density of $3\times10^{4}$\cmm3 is reached. This increase coincides with the heating rate being dominated by cosmic ray heating ($<10^4$ \cmm3) and compressional heating ($>10^4$\cmm3) while photo-electric heating has decreased strongly at these densities due to dust shielding. As the temperature is rising in the surface chemistry model up to 30\K, CO ice starts to evaporate back into the gas phase, resupplying the cloud with efficient coolants. At the same time, gas-grain coupling grows stronger with increasing density and becomes the dominant coolant above \nhtot = $3\times10^{4}$\cmm3. Gas-grain collisional cooling is the main mechanism that drives the temperature to decline again in the surface chemistry model. Above $3\times10^5$\cmm3 both models see their temperatures increase once again due to compressional heating.

The temperature of the model without surface chemistry drops to 8\K and lower, but rises above 10\K at \nhtot = $10^4$\cmm3. The general shape of this model is in line with \cite{2005ApJ...626..627O}.

\subsection{CO depletion}
\label{sec:codepletion}
CO is an important coolant in molecular clouds. Especially below temperatures of 20\K and at densities of 10$^3$\cmm3 or higher. CO rotational transition cooling dominates the thermal balance in this regime if the CO abundance is higher than 10$^{-5}$. Without freeze-out, the CO abundance is nearly at its maximum of 1.3$\times 10^{-4}$ at densities of \nhtot $\geq$ 10$^3$\cmm3. If we do take freeze-out into account, this molecule is depleted from the gas phase at low, $\lesssim 20$\K temperatures. While the accretion rate of dust grains scales with the gas temperature as $\sqrt{T_{\rm gas}}$, the evaporation rate follows the dust temperature exponentially as $\exp(\rm -E_{B}/T_{\rm dust})$. E$_{\rm B}$ is here the binding energy of the specific species with a value of E$_{\rm B} = 830$\K for CO on bare grains \citep{2003Ap&SS.285..633C,2012MNRAS.421..768N}.

\begin{figure}
\includegraphics[scale=0.455]{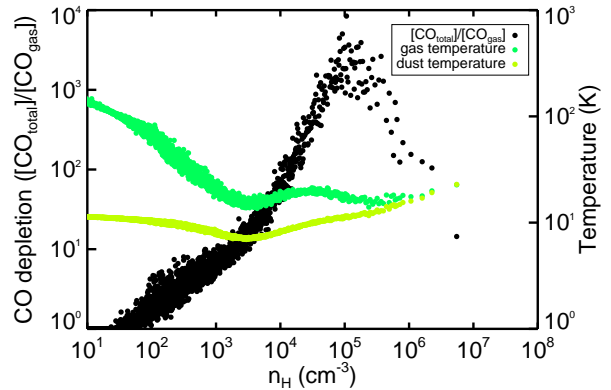}
\caption{CO depletion. The CO$_{\rm gas+ice}$ over CO$_{\rm gas}$ ratio is plotted against \nhtot at $t = t_{\rm ff}$. The depletion factor is given as black dots with their values along the Y-axis on the left. The gas and dust temperatures are given as green and yellow dots with their values along the Y-axis on the right.}
\label{fig:depletion}
\end{figure}

In Fig. \ref{fig:depletion}, we plot the CO depletion factor, which we define as the ratio of the total CO abundance ($\rm CO_{ice} + CO_{gas}$) over the CO abundance in the gas phase ($\rm CO_{gas}$), against number density.
The gas and dust temperatures are overplotted in this figure with their values on the right Y-axis. As expected, CO depletion is correlated with the temperatures. As the dust temperature decreases, the CO molecule is significantly depleted. The depletion is also aided by the increase in number density. This deprives the cloud of its main coolant and heating starts to dominate. Once the gas temperature starts to increase, so does the dust temperature due to collisional coupling, and CO evaporates back into the gas phase. The depletion drops to the point where CO is mostly in the gas phase when both temperatures are above 25\K at densities of $\gtrsim$10$^{7}$\cmm3. There is a slight delay between the temperature change and the change in CO depletion, but this can be explained by the time dependent nature of the chemical rate equations and the thermal balance.

\subsection{Heating and cooling rates}
\label{sec:heatingandcooling}
\begin{figure}
\includegraphics[scale=0.455]{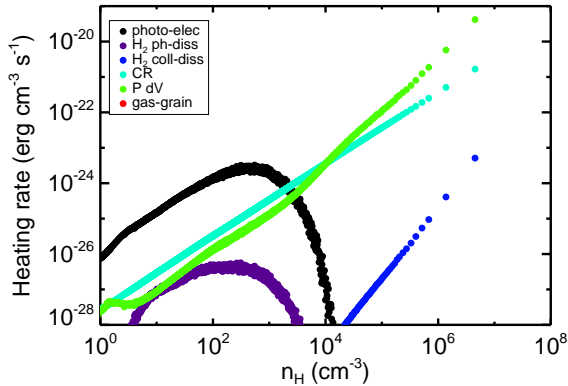}
\caption{Heating rates as a function of \nhtot. The heating rates for the two models are similar, therefore, only the model with GSC is shown at $t = t_{\rm ff}$.}
\label{fig:heatingrates}
\end{figure}

By affecting the chemistry and changing the molecular abundances in the gas phase, dust grains have an indirect effect on the thermal balance, aside from the direct effect that they have due to gas-grain coupling. The heating rates are not influenced, because the main heating mechanisms do not directly depend on molecular abundances. We show in Fig. \ref{fig:heatingrates} the heating rates as a function of number density.
We can see in this figure that photo-electric heating dominates until $2\times10^3$\cmm3, cosmic ray heating between $2\times10^3$ and 10$^4$\cmm3, and compressional heating above 10$^4$\cmm3.

\begin{figure}
\includegraphics[scale=0.455]{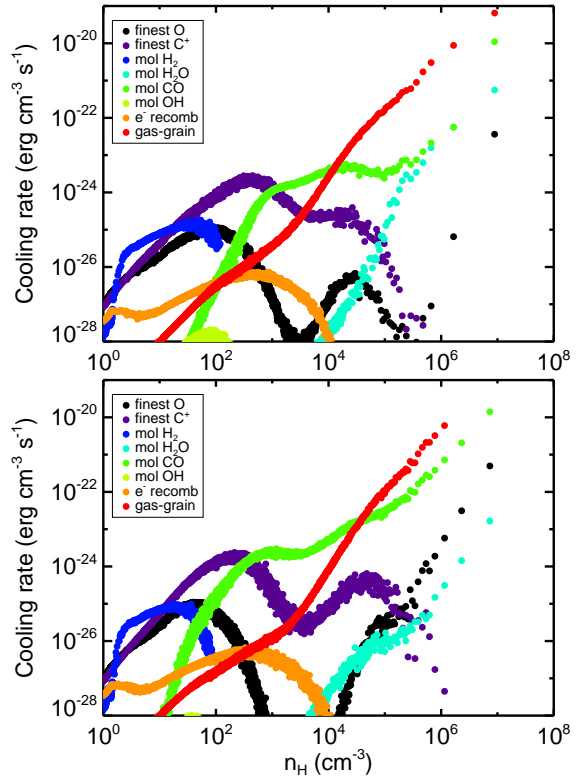}
\caption{Cooling rates as a function of \nhtot. In the top panel, the cooling rates of the model with GSC is shown. In the bottom panel, the cooling rates of the model without GSC is shown. The images are displayed at $t = t_{\rm ff}$.}
\label{fig:coolingrates}
\end{figure}

Cooling, on the other hand, is affected by atomic and molecular abundances. Molecules such as \hh, \h2o, \co, \oh, but also important atoms like \cz and O are key players in determining the cooling rate at various densities and temperatures. In Fig. \ref{fig:coolingrates}, we show the cooling rates as a function of number density for the two models.
In this case, the impact of surface chemistry is significant. The cooling rates that are obtained from ro-vibrational transitions of \co, \h2o, and the fine-structure lines of \cz and O are notably altered. The cooling due to gas-grain coupling has also changed, which is higher with surface chemistry, but this is due to the increased gas temperature in the surface chemistry model. Of these, the most important coolant to influence the thermal balance is \co. Where in the pure gas chemistry model CO becomes the dominant coolant over \cz at a density of \nhtot = $0.40\times10^3$\cmm3, the surface chemistry model achieves this at \nhtot = $1.25\times10^3$\cmm3. The difference becomes more pertinent around \nhtot = $10^4$\cmm3 where the cooling rates are on the order of $10^{-23}$\ergcms. This drop in cooling due to freeze-out causes the higher temperatures in the model with surface chemistry. At high densities, \nhtot $\gtrsim 3\times10^4$\cmm3, gas-grain collisional cooling dominates the gas temperature until compressional heating takes over above \nhtot = $3\times10^5$\cmm3, which causes the models to converge and evolve similarly in temperature from this point on.

\subsection{Equation of state}
\label{sec:eos}
\begin{figure}
\includegraphics[scale=0.455]{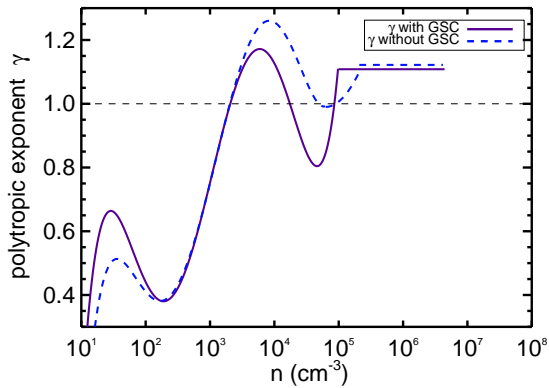}
\caption{Polytropic exponent $\gamma$. The solid line depicts the results of the model with grain surface chemistry. The dashed line depicts the results of the model without grain surface chemistry.}
\label{fig:eos}
\end{figure}

In our hydrodynamical code we employed a polytropic equation of state, $\rm P \propto \rho^{\gamma}$ \citep{2000ApJ...538..115S}, where $\gamma$ is the polytropic exponent. For an ideal gas equation of state, $\rm P \propto \rho T_g$, the polytropic exponent is written as
\begin{equation}
\gamma = \frac{d{\rm log}P}{d{\rm log}\rho} = 1 + \frac{d{\rm log}T_g}{d{\rm log}n},
\end{equation}
where $n = \rho/\mu m_h$, with $\mu$ as the mean molecular weight. Note that this differs from $n_H$. From the derivative of the temperature profile function (Fig. \ref{fig:phases}), as a function of $n$ instead of \nhtot, we can retreive the EOS. In order to obtain this we fit a line through our temperature profile by a two part fit; a higher order polynomial function that encompasses the range from \nhtot $\simeq$ 1 to $10^5$\cmm3 and a power-law function for \nhtot $\gtrsim 10^5$\cmm3. For a collapsing molecular cloud with detailed grain surface chemistry and freeze-out $\gamma$ is given by
\begin{eqnarray}
\gamma_{n<10^5} &=& 0.1158 x^5 - 1.7620 x^4 + 10.1195 x^3 - 27.1393 x^2 +
\nonumber \\
 && 33.8331 x - 15.1377
\nonumber \\
\gamma_{n>10^5} &=& 1.1083,
\label{eq:eq1}
\end{eqnarray}
where $x = {\rm log} \,n$. We expect $\gamma$ to follow the established power-law until densities of \nhtot $\simeq$ 10$^{11}$ \cmm3 \citep{2005ApJ...626..627O, 2010ApJ...722.1793O}. We show a plot of the polytropic exponent $\gamma$ in Fig. \ref{fig:eos}.

\section{Conclusions and discussion}
\label{sec:conclusion}
In our model with surface chemistry, we find that the depletion of coolants from the gas phase, mainly \co, has an impact on the thermal evolution, especially in the density range of $10^3 - 10^5$\cmm3. The \co molecules are mostly frozen on dust grains at cloud temperatures of below 20\K. This raises the temperature of the cloud locally and causes a thermal oscillation between 20-30\K as \co starts to evaporate back into the gas phase again above $\sim$25\K. At a certain point, and when \nhtot $\simeq 3\times10^4$\cmm3, collisional cooling on dust grains becomes the dominant channel for both models. This balances the gas temperature and couples it to the dust temperature. We see only little difference in the thermal evolution between the two models above $3\times10^5$\cmm3.

We also present the EOS of a collapsing molecular cloud with grain surface chemistry and freeze-out included in the models (see Eq. \ref{eq:eq1}). The EOS is softer, $\gamma$ drops far below unity, in the model with grain surface chemistry between densities of $2\times10^3 - 10^5$\cmm3. This is expected to have consequences for molecular cloud fragmentation and the distribution of stellar masses \citep{1985MNRAS.214..379L, 2000ApJ...538..115S, 2003ApJ...592..975L}. The higher Jeans mass in the GSC model due to higher gas temperatures, combined with the higher compressibility of the gas following a drop in $\gamma$ back to below unity, is expected to result in higher mass cores with more massive stars forming inside them compared to the pure gas phase model. This should lead to a larger charachteristic mass of the GSC initial mass function.

We conclude that surface chemistry on dust grains significantly affects the thermodynamics and molecular abundances of molecular clouds. Moreover, we note that these changes impact the early stages of molecular clouds. We intend to further study and discuss the details of freeze-out on cloud fragmentation, with higher resolution simulations, including the formation of stars as sink particles, in a follow-up paper.

\section*{Acknowledgments}
The software used in this work was developed in part by the DOE NNSA ASC- and DOE Office of Science ASCR-supported Flash Center for Computational Science at the University of Chicago. The simulations have been run on the dedicated special purpose machines `Gemini' at the Kapteyn Astronomical Institute, University of Groningen and at the Donald Smits Center for Information Technology (CIT) using the Millipede Cluster, University of Groningen. Some of the kinetic data used in this work has been downloaded from the online database \textit{KiDA} (\cite{2012ApJS..199...21W}, http://kida.obs.u-bordeaux1.fr).

\bibliography{biblio.fifth.bib}

\end{document}